\documentclass[]{aipproc}
\layoutstyle{6x9}
\SetInternalRegister\hbadness{8000} 
\newcommand{\gsim}{\raisebox{-0.07cm}{$\, \stackrel{>}{{\scriptstyle
\sim}}\, $}}
\begin{document}
\noindent
DESY 01--107 \hfill {\tt hep-ph/0107317}\\
July 2001

\title 
      [QCD Analysis of Polarized Deep Inelastic Scattering Data]
{QCD Analysis of Polarized Deep Inelastic Scattering Data}


\author{J. Bl\"umlein}{
  address={DESY Zeuthen, Platanenallee 6, 15738 Zeuthen, Germany},
  email={johannes.bluemlein@desy.de},
}

\iftrue
\author{H. B\"ottcher}{
  address={DESY Zeuthen, Platanenallee 6, 15738 Zeuthen, Germany},
  email={helmut.boettcher@desy.de},
}
\fi

\copyrightyear  {2001}

\begin{abstract}
A QCD analysis of the world data on inclusive polarized deep inelastic 
scattering of leptons on nucleons is presented in leading and 
next--to--leading order. New 
parameterizations are derived for the quark and gluon distributions and the 
value of $\alpha_s(M_Z)$ is determined. Emphasis is put on the derivation 
of fully correlated error bands for these distributions which are directly 
applicable to determine experimental errors of other polarized observables. 
The impact of the variation of both the renormalization and factorization 
scales on the value of $\alpha_s$ is studied. Finally a factorization--scheme 
invariant QCD analysis based on the observables $g_1(x,Q^2)$ and 
$d g_1(x,Q^2)/d \log(Q^2)$ is performed in next--to--leading order, which is 
compared to the standard analysis.
\end{abstract}

\date{\today}

\maketitle

\section{Introduction}

The remarkable growth of experimental data on inclusive polarized deep 
inelastic scattering of leptons on nucleons over the last years 
[1--9]~allows to perform
refined QCD analyses of polarized structure functions in order to reveal 
the spin--dependent partonic structure of the nucleon. A number of 
such analyses has already been worked out. The most recent ones are
[10--13]~\footnote{For a more complete list of references see the references 
therein and Ref.~[14]}. In this talk results from a new QCD analysis in 
leading (LO) and next--to--leading (NLO) order [14]~\footnote{All details of 
the analysis are given in Ref.~[14].} are presented. New parameterizations of 
the polarized quark and gluon distributions are derived including the 
parameterizations of fully 
correlated $1\sigma$ error bands for these distributions, which are directly 
applicable to calculate errors of other polarized observables. Furthermore 
the value of $\alpha_s(M_Z)$ is determined. Finally and for the first time a
factorization--scheme independent QCD evolution based on the observables 
$g_1(x,Q^2)$ and $d g_1(x,Q^2)/d \log(Q^2)$ in next--to--leading order is 
performed.

\section{Formalism}

In LO the polarized structure function $g_1(x,Q^2)$ is expressed as the 
sum of the polarized quark distributions $\Delta q_i(x,Q^2)$ weighted by 
the square of the quark charges.
In NLO the expression for $g_1(x,Q^2)$ involves the polarized singlet 
$\Delta \Sigma(x,Q^2)$, the gluon $\Delta G(x,Q^2)$, and the non--singlet 
$\Delta q^{NS}(x,Q^2)$ distributions and reads

\begin{equation}
g_1(x,Q^2) = \frac{1}{2} \Biggl[
\left( \frac{1}{n_f} \sum\nolimits_{i=1}^{n_f}
e_i^2 \right) \left[ \delta C_S \otimes \Delta \Sigma  + 
                     \delta C_G \otimes \Delta G \right] +
                     \delta C_{NS} \otimes \Delta q^{NS} \Biggr],
\end{equation}

\noindent
where $n_f$ is the number of active quark flavors and $e_i$ is the quark 
charge. The symbol $\otimes$ denotes the Mellin
convolution w.r.t. $x$ of the polarized 
parton densities $\Delta q_i(x,Q^2)$ with the corresponding polarized Wilson 
coefficient functions $\delta C_i(x,\alpha_s(Q^2))$. The polarized singlet 
and non--singlet distributions are certain combinations of the polarized quark 
distributions $\Delta q_i(x,Q^2)$. 

The evolution equations used to evolve the parton densities to different 
$Q^2$ values contain the polarized splitting functions 
$\Delta P_{ij}(x,\alpha_s(Q^2))$. Both the polarized Wilson coefficient 
[15] and the polarized splitting functions [16] are known in
the $\overline{MS}$ scheme up to order ${\cal O}(\alpha_s^2)$. 

\section{Method}

The shape chosen for the
parameterization of the polarized parton distributions at the input 
scale of $Q^2 = 4.0~GeV^2$ is~:

\begin{equation}
x\Delta q_i(x,Q_0^2) = \eta_i A_i x^{a_i} (1 - x)^{b_i} 
(1 + \gamma_i x + \rho_i x^{\frac{1}{2}}).
\end{equation}

\noindent
The normalization constant $A_i$ is chosen such that $\eta_i$
is the first moment of $\Delta q_i(x,Q_0^2)$.
The densities to be fitted are $\Delta u_v$\footnote{Note that: $\Delta q + \Delta \bar{q} = \Delta q_v + 2\Delta \bar{q}$.},
$\Delta d_v$, $\Delta \bar{q}$, and $\Delta G$.

Assuming $SU(3)$ flavor symmetry the first moments of $\Delta u_v$ and  
$\Delta d_v$ are determined by the $SU(3)$ parameters $F$ and $D$ 
measured in neutron and hyperon $\beta$--decays and can be fixed to 
$\eta_{u_v} = 0.926$ and $\eta_{d_v} = -0.341$. In addition we assume a 
flavor symmetric sea, i.e. only one general sea distribution 
$\Delta \bar{q}(x,Q^2)$ is required. No assumptions are made concerning 
positivity and helicity retention. Given the present accuracy of the data 
we set a number of parameters to zero, namely $\rho_{u_v} = \rho_{d_v} = 0$, 
$\gamma_{\bar{q}} = \rho_{\bar{q}} = 0$, and  $\gamma_G = \rho_G = 0$. 
This choice reduces the number of parameters to be fitted for each parton 
distribution to three. In addition the parameter $\Lambda_{QCD}$ was 
determined. The relative normalizations of the different 
data sets were fitted and then fixed. Doing so part of the experimental 
systematics was taken into account.

\section{Results}

\begin{figure}[htb]
\includegraphics[height=.60\textheight]{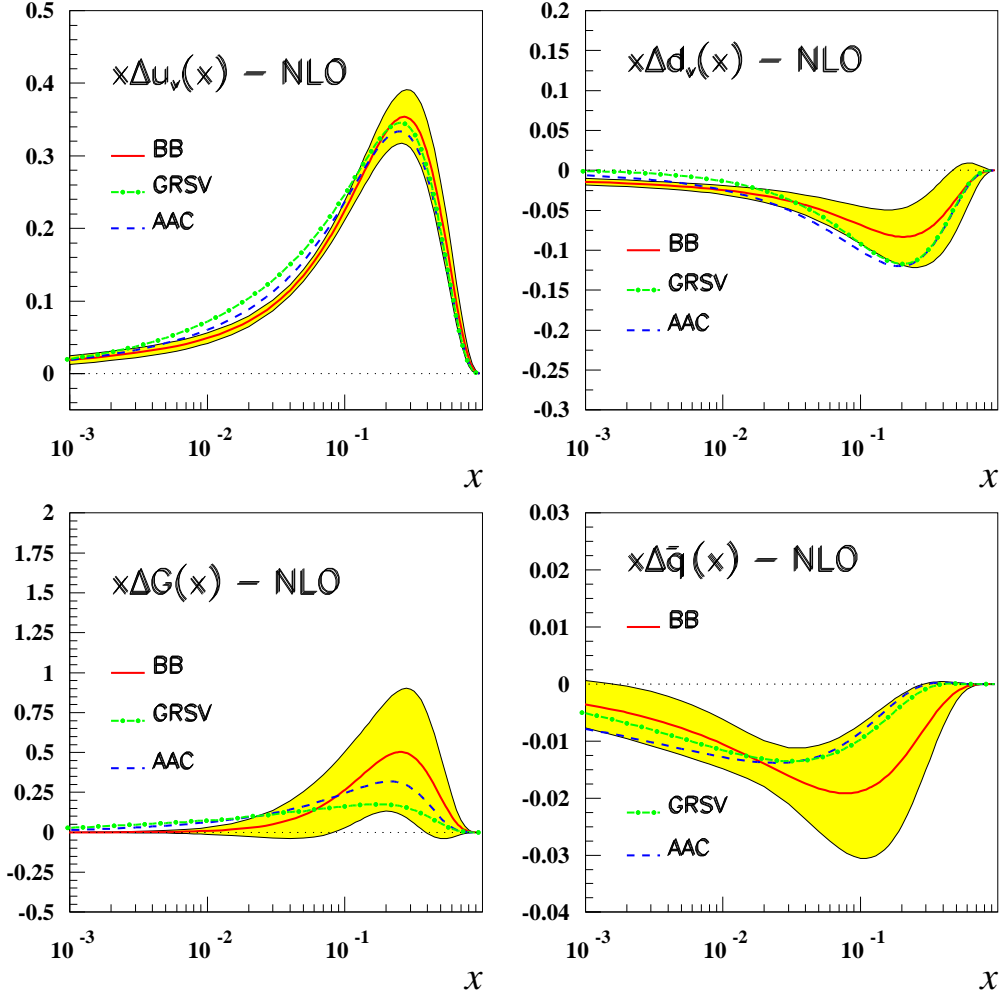}
\caption{\label{deltaq}
Polarized parton distribution at the input scale $Q_0^2 = 4.0~ GeV^2$ 
(solid line) compared to results obtained by GRSV (dashed--dotted line) [12] 
and AAC (dashed line) [10]. The shaded areas
represent the fully correlated $1\sigma$ error bands calculated by 
Gaussian error propagation, Ref.~[14].}
\end{figure}

The results reported here are based on 433 data points of asymmetry data, 
i.e. $g_1/F_1$ or $A_1$, above $Q^2 = 1.0~GeV^2$, the world statistics 
published so far. The QCD fits are performed on $g_1$ which is evaluated 
from the asymmetry data using parameterizations for the unpolarized structure 
functions $F_2$ [17] and $R$ [18]. We realized that the 4 parameters 
$\gamma_{u_v}$, $\gamma_{d_v}$, $b_{\bar q}$, and $b_G$ had to be fixed in 
addition at their values at $\chi^2_{\rm min}$ since 
the data do not constrain these parameters well enough. Only fits with a 
positive definite covariance matrix were accepted in order to be able to 
calculate the fully correlated 1$\sigma$ error bands. The NLO polarized parton 
densities at the input scale are presented in Fig.~\ref{deltaq}. 

\begin{figure}[htb]
\includegraphics[height=.60\textheight]{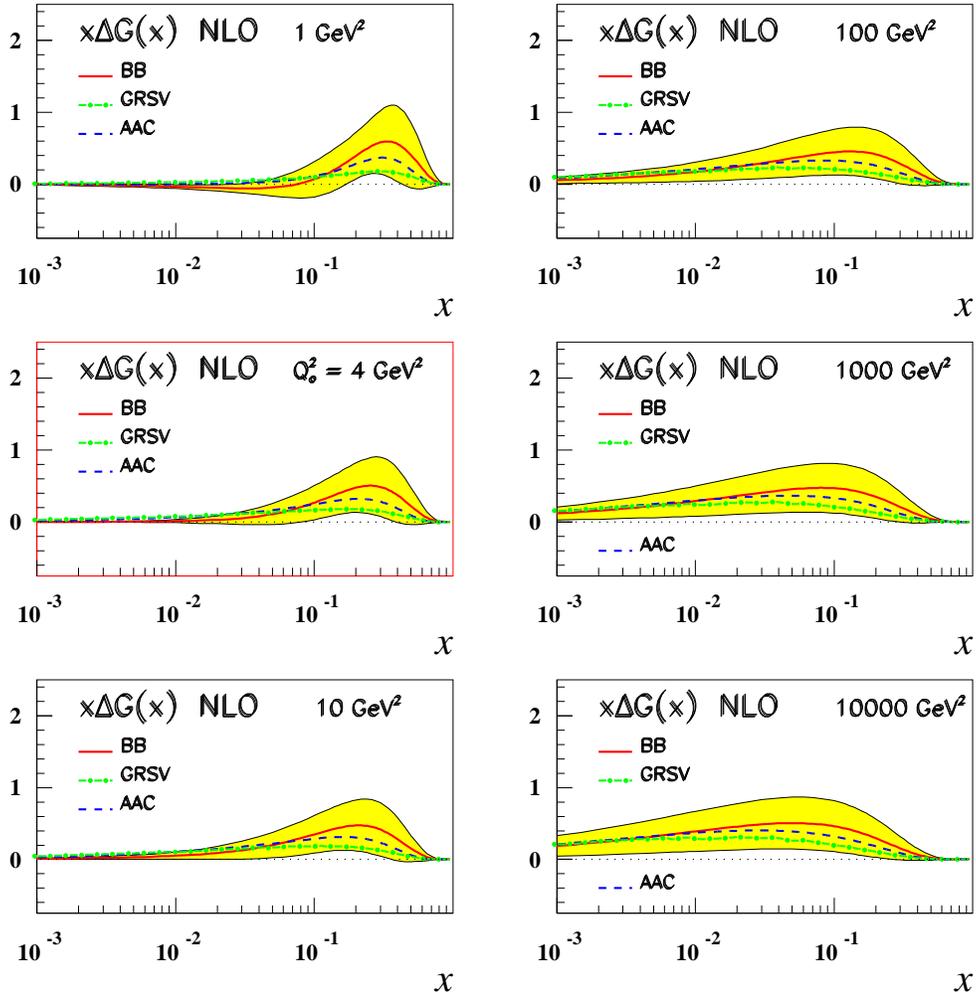}
\caption{\label{evolq}The polarized parton distribution $\Delta G$ evolved 
up to $Q^2$ values up to $Q^2 - 10,000~Gev^2$ (solid line) compared to 
results obtained by GRSV (dashed--dotted line) [12] and AAC (dashed
line) [10].
The shaded areas represent the fully correlated $1\sigma$ error bands 
calculated by Gaussian error propagation, Ref.~[14].}
\end{figure}

While the quality of the data is sufficient to determine $\Delta u_v$ and 
$\Delta d_v$ with good accuracy, $\Delta G$ and $\Delta \bar q$ have much 
broader error bands. This is essentially due to the lack of data at low $x$. 
The agreement with the results of the analyses of 
Refs. [10] and [12] is satisfactory within the error bands.
The measured structure function $g_1^p$ is well described both as 
function of $x$ and of $Q^2$. The derived parton distributions and its error 
bands have been evolved to $Q^2$ values up to $10,000~GeV^2$. As an example 
the evolution of $\Delta G$ is shown in Fig.~\ref{evolq}. One observes that 
even within the error band $\Delta G$ stays {\it positive} up to the highest 
$Q^2$ value. It should be mentioned that $\Delta \bar q$ develops a trend 
to change sign and becomes slightly positive towards higher 
$Q^2$ values and for $x \gsim 0.1$ within the errors.

In determining $\alpha_s(M_z^2)$ the parameter $\Lambda_{QCD}$ was fitted. 
The impact of the variation of both the renormalization and factorization 
scales on the value of $\alpha_s$ was studied. The following value for 
$\Lambda_{\rm QCD}$ was obtained 

\begin{eqnarray}
 \Lambda_{\rm QCD}^{(4)}  =  241 \pm 58~{\rm (fit)}
\begin{array}{l} + 65 \\ -44 \end{array}{\rm (fac)}
\begin{array}{l} + 117 \\ -58 \end{array}{\rm (ren)}  \quad MeV, \nonumber
\end{eqnarray}

\noindent
which results into a value of 

\begin{eqnarray}
 \alpha_s(M_Z^2)  =  0.114
\begin{array}{l} + 0.004\\-0.005 \end{array}{\rm (fit)}
\begin{array}{l} + 0.005\\-0.004 \end{array}{\rm (fac)}
\begin{array}{l} + 0.008\\-0.005 \end{array}{\rm (ren)}. \nonumber
\end{eqnarray}

\noindent
This value of $\alpha_s(M_Z^2)$ is compatible
within the errors  with the world average of $0.118 \pm 0.002$ [19] 
and with values from other QCD analyses [20], although the central value 
tends to be lower, as also in Ref. [20b].
 
\begin{figure}[htb]
\includegraphics[height=.60\textheight]{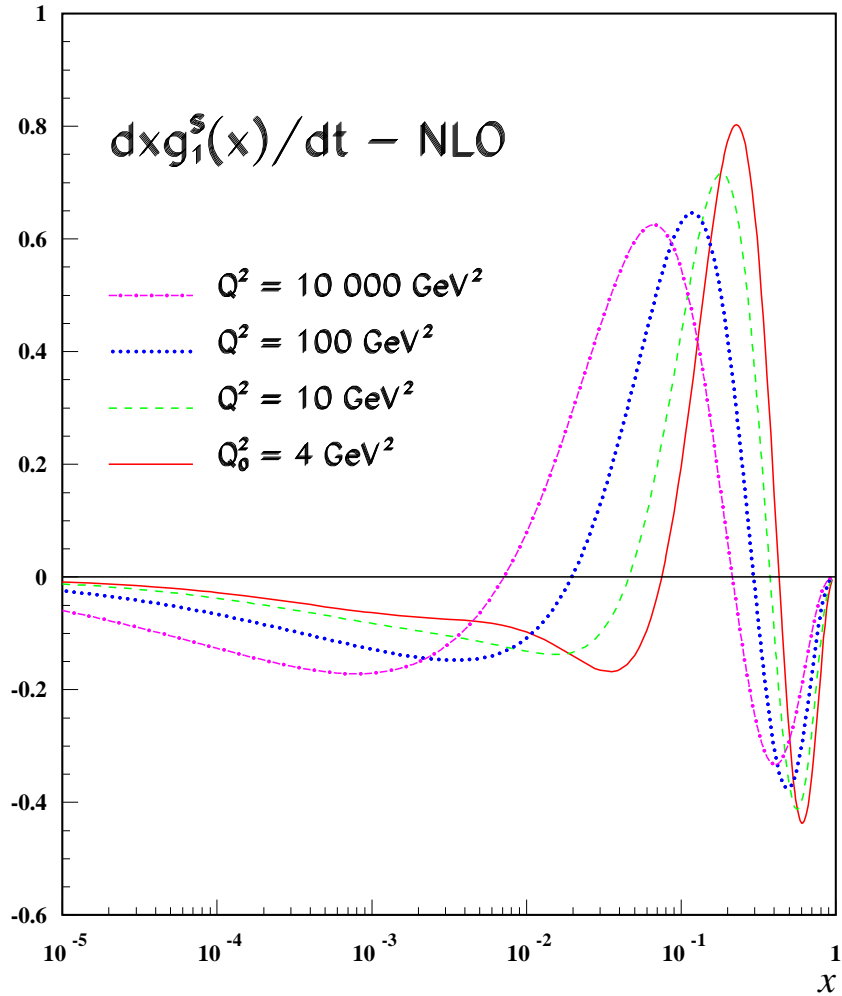}
\caption{\label{slopeg1}The evolution of $d g_1^s(x,Q^2)/d t$ (singlet 
contribution) with
$t=-2/\beta_0 \ln(\alpha_s(Q^2)/\alpha_s(Q_0^2))$. The
slope of $g_1$ was determined from a fitted $g_1$, see text.}
\end{figure}

Finally  a factorization--scheme invariant QCD analysis 
based on the observables $g_1(x,Q^2)$ and $d g_1(x,Q^2)/d \log(Q^2)$ in 
next-to-leading order was performed. The corresponding evolution equations
have been  worked out in Ref.~[21]~\footnote{The same case has already
been considered in Ref.~[22].}. Such an analysis has the advantage of
direct control over the input since it comes from measured quantities. 
The only parameter to be determined is $\Lambda_{QCD}$. 
Unfortunately, the present data do not yet allow to determine the slope  
$\partial g_1(x,Q^2)/\partial \log(Q^2)$ as an input density from 
measurements of $g_1$, but it is derived here from the fit result for $g_1$ 
described above. The evolution of the so determined slope is shown in 
Fig.~\ref{slopeg1}.

\noindent
A downward 
shift of $12~MeV$ in $\Lambda_{QCD}$ was found yielding a similar result 
for $\alpha_s(M_Z^2)$ as obtained in the standard analysis.

\section{Conclusions}

An LO and NLO QCD Analysis of the current World--Data on Polarized Structure 
Func\-tions was performed. New parameterizations of the polarized parton 
densities including their errors were derived. They are available via a fast 
{\tt FORTRAN} code for the range: $1 < Q^2 < 10^6~{\rm GeV}^2$ and
$10^{-4} < x < 1$. The value determined for $\alpha_s(M_Z^2)$ is compatible 
with the world average, although the central value obtained is lower.
First steps in a factorization--scheme invariant QCD evolution based on the 
structure function $g_1(x,Q^2)$ and 
$\partial g_1(x,Q^2) / \partial \log Q^2$ were performed yielding similar 
results for $\alpha_s(M_Z^2)$. This latter analysis is a very promising way 
to proceed in the future, since it allows to extract $\Lambda_{\rm QCD}$ 
fixing all the input distributions by direct measurements.

\begin{theacknowledgments}
This work was supported in part by EU contract FMRX-CT98-0194 (DG 12 - MIHT). 
For discussions in an early phase of this work we would like to thank A.~Vogt.
\end{theacknowledgments}

\vspace*{-2mm}

\end{document}